\documentclass{emulateapj}

\slugcomment{Submitted to ApJ Letters December 18, 2006; Accepted April 04, 2007}
\shorttitle{Extreme Debris Disk Asymmetries}
\shortauthors{Kalas, Fitzgerald, Graham.}

\begin{document}

\title{Discovery of extreme asymmetry in the debris disk surrounding HD 15115}

\author{Paul Kalas\altaffilmark{1,2}, Michael P.
Fitzgerald\altaffilmark{1,2}, James R. Graham\altaffilmark{1,2}}
\affil{}

\altaffiltext{1}{Astronomy Department and Radio Astronomy Laboratory, 
601 Campbell Hall, Berkeley, CA 94720}
\altaffiltext{2}{National Science Foundation Center for Adaptive Optics, University of California, Santa Cruz, CA 95064}

\begin{abstract}
We report the first scattered light detection of a dusty debris disk surrounding the F2V star
HD 15115 using the {\it Hubble Space Telescope} in the optical, and Keck adaptive
optics in the near-infrared.  The most remarkable property of the  HD 15115 disk relative
to other debris disks is its extreme length asymmetry.
The east side of the disk is detected to $\sim$315 AU radius, whereas the west side of the
disk has radius $>$550 AU.  We find a blue optical to near-infrared scattered light color relative
to the star  that indicates grain scattering properties similar to the AU Mic debris disk.  
The existence of a large debris disk surrounding HD 15115 adds further evidence for membership 
in the $\beta$ Pic moving group, which was previously argued based
on kinematics alone.  Here we hypothesize that the extreme disk asymmetry is due to 
dynamical perturbations from HIP 12545, an M star 0.5$\degr$ (0.38 pc) east of HD 15115 
that shares  a common proper motion vector, heliocentric distance, galactic space velocity,
and age.  
\end{abstract}

\section{Introduction}
Volume-limited, far-infrared surveys of the solar neighborhood suggest that $\sim$15\% of
main sequence stars have excess thermal emission indicative of circumstellar grains \citep{aumann85, bap93, meyer07}.  
Direct imaging of dust scattered light reveals the geometry of the grain population relative
to the star, which further elucidates the origin of dust.  In some cases, a circumstellar nebulosity may be 
amorphous with asymmetric striated features  produced when stellar radiation pressure deflects
interstellar dust \citep{kalas02}.  In other cases, such as $\beta$ Pictoris and Fomalhaut, the 
geometry of dust is consistent with a circumstellar disk or belt related to the formation of 
planetesimals \citep{smith84, kalas05}.   Though larger bodies such as comets and asteroids
are not directly observed, they most likely exist as a reservoir for injecting fresh debris
into the system over the lifetime of the star.  Furthermore, circumstellar debris disks display
significant structure and asymmetry that may be linked, in principle, to dynamical perturbations from a
planetary system \citep{roques94, liou99, moro02}.   Unfortunately, only $\sim$10\% of stars with excess thermal emission
have detected scattered light disks due to the high contrast between the host star
and the low surface brightness nebulosity at optical and near-infrared wavelengths.   
Fortunately, the observational capabilities
have improved in recent years due to instrument upgrades on the Hubble Space
Telescope (HST) and the implementation of adaptive optics (AO) on large, ground-based
telescopes.  

Here we show new scattered light images of a debris disk surrounding HD 15115, an F2 star at 45 pc (Table 1), 
first reported as a source of thermal excess emission by \citet{silverstone00}.  The spectral energy 
distribution is consistent with a single temperature dust
belt at $\sim$35 AU radius with an estimated dust mass of 0.047 M$_{\earth}$ \citep{zuck04, williams06}.
Recently, \citet{moor06} identified HD 15115 as a candidate for membership
in the 12 Myr-old $\beta$ Pic moving group (BPMG), based on new radial velocity measurements
that resulted in galactic kinematics similar to those of the BPMG.  

\section{Observations \& Data Analysis}
We first detected the HD 15115 disk in scattered light using the HST ACS High Resolution Camera (HRC)
on 2006 July 17.  We used the F606W broadband filter  and the 1.8$\arcsec$ diameter occulting
spot to artificially eclipse the star.
Three flatfielded frames of 700 seconds each from standard pipeline processing of the HST data archive 
were median combined for cosmic ray rejection.
The point spread function was then subtracted iteratively using five other stars of similar
spectral type obtained from the HST archive.  The relative intensity scaling between images was iteratively 
adjusted until the residual image showed a mean radial profile equal to zero 
intensity perpendicular to the circumstellar disk.  
The images were then corrected for geometric distortion, giving a 25 mas pixel$^{-1}$ scale.   

The resulting optical images revealed a needle-like feature projecting westward from the star to the
edge of the field, but with almost no counterpart to the east.   Given the high degree of asymmetry
that could conceivably arise from instrumental scattering, we endeavored to confirm the disk
using the Keck II telescope with AO on 2006 October 07  and  2007 January 26.  Utilizing the near-infrared
camera NIRC2,  a 0.4$\arcsec$ radius occulting spot and a 10 mas pixel scale, we confirmed the
existence of the disk in J (1.2 $\mu$m), H (1.6 $\mu$m), and K$^\prime$ (2.2 $\mu$m).  PSF subtraction
is accomplished by allowing the sky to rotate relative to the detector, thereby separating the stellar
PSF from the disk.  
The observing procedure and data reduction procedure are fully described in 
\citet{fitz07}.  Due to poor observing conditions in October, including intermittent cirrus clouds, we used only
the best fraction of data by visually selecting frames of relatively constant intensity and PSF sharpness.  
The resulting effective integration times
are 450 s, 980 s, and 600 s for J, H, and K$^\prime$, respectively.   Standard star observations
were obtained under similar, non-photometric conditions and processed in a similar manner.  
In January 2007 we re-observed HD 15115 (1930 s cumulative integration time) and two standard
stars under photometric conditions from Keck using the same instrumentation with the H broadband filter.  
However, the observations were made after meridian transit 
and the limited rotation of the sky relative to the instrumental PSF causes disk emission to be included 
in the PSF estimate, resulting in disk self-subtraction at small radii.  Our analysis of the 2007 January
data therefore yields a detection of the west ansa in the region 1.3$\arcsec - 3.3\arcsec$ radius.
The photometry in this second epoch agrees well with that of the first epoch (on average, the 
2007 January disk photometry is 0.13 mag arcsec$^{-2}$ fainter than 2006 October), suggesting that our
frame selection technique for the first epoch of cloudy conditions effectively filtered out non-photometric data.

\section{Results}
Fig. 1 shows the PSF-subtracted images of HD 15115 with HST and with Keck.
The west side of the disk in the optical HST data has PA $= 278\degr.5 \pm 0.5$ and
is detected from the edge of the occulting spot at 1.5$\arcsec$ (67 AU) radius
to the edge of the field at 12.38$\arcsec$ (554 AU) radius.  The east midplane
is detected as far as $\sim$7$\arcsec$ (315 AU) radius.  At this radius the east midplane
begins to intersect the outer portion of the coronagraph's 3.0$\arcsec$ occulting spot.  
Further east, past the spot and to the edge of the field, no nebulosity is detected 9.0$\arcsec - 14.9\arcsec$ radius.
The appearance of the disk is more symmetric in the 2006 October Keck data, which 
show the disk between 0.7$\arcsec$ (31 AU) - 2.5$\arcsec$ (112 AU).  

Optical surface brightness contours (Fig. 2) reveal a sharp midplane morphology for the
west extension that indicates an edge-on orientation to the line of sight.  The west midplane is qualitatively
similar to that of $\beta$ Pic's northeast midplane, including a characteristic
width asymmetry \citep{kalas95, goli06}.  The northern side of the west midplane is more vertically
extended than the southern side.  For example, the full-width at half-maximum
across the disk midplane at 2$\arcsec$ radius is 0.19$\arcsec \pm 0.10\arcsec$ 
in both the optical and NIR data.  However, the vertical cuts are not
symmetric about the midplane when measuring the half-width at quarter-maximum (HWQM).
The HWQM north of the west midplane is $1.6\pm 0.1$ times greater than that of the HWQM south of the west midplane.
This width asymmetry is
confirmed in the Keck data.  If the width asymmetry is found to be in the opposite direction in
the opposite midplane, then \citet{kalas95} refer to such a feature as the butterfly asymmetry.
The butterfly asymmetry is evident in the morphology of $\beta$ Pic, that \citet{goli06} recently related
to the presence of a second disk midplane tilted relative to the main disk midplane.
However, our detection of HD 15115's east midplane has insufficient signal
to noise to confirm the presence of a width asymmetry here.

We note that none of the surface brightness
profiles show evidence for significant flattening inward toward the star (Fig. 3).  
All four surface brightness profiles
are well-represented by a single power law decrease
with radius.  If there is an inner dust depletion, then it resides within 40 AU radius.
This constraint is consistent with model fits of the spectral energy distribution
that place the dominant emitting dust component at $\sim$35 AU radius
\citep{zuck04, williams06}.

The color of the disk may be estimated in the $2.0\arcsec - 3.3 \arcsec$ region
where the H-band and V-band data overlap (Fig. 3).  At face value, $\Sigma_V - \Sigma_H \approx -0.6$ mag arcsec$^{-2}$
at 2$\arcsec$ radius, increasing to $-1.9$ mag arcsec$^{-2}$ at 3.3$\arcsec$ radius for the West disk
extension.  The east ansa has similar blue scattering at 2$\arcsec$ radius,
but the V-band surface brightness profile is steeper in the east than in the west, giving a roughly
constant blue color with increasing radius in the east. 

In a future paper we will present a detailed model of dust scattering and thermal emission properties,
which requires a more complicated treatment of the obvious disk asymmetry.
However, for isotropically scattering grains in an edge-on disk, an analytic approach
shows that the grain number density distribution as a function of radius within the disk
midplane follows a power-law with index equivalent to one
minus the sky-projected radial midplane power-law index.  From the Keck data in Fig. 3, we estimate that
the disk number density distribution decreases with disk radius as $r^{-3}$ in the inner region
up to $\sim$3.3$\arcsec$ radius for both sides of the disk.  At $>3.3\arcsec$ radius, the optical data show that
this profile continues for the east extension, but that
the disk number density profiles flattens for the west extension, 
as described in Fig. 3.  A precise measurement of the color and polarization of the disk scattered light is
necessary to further constrain the grain size distribution, the corresponding scattering phase
function and albedo, and the effect on the disk number density profile.  

\section{Discussion}
Asymmetric disk structure is evident in the majority of debris disks,
and most authors invoke planetary perturbations as the likely origin.
Secular perturbations may offset the center of global disk symmetry from the 
location of the star, though this effect may also be produced by
an external perturber \citep{wyatt99}.  The edge-on debris disk surrounding $\beta$
Pic displays a variety of radial and vertical asymmetries on large scales \citep{kalas95}
that may be most relevant to the study of HD 15115.  In the deepest optical images
of the $\beta$ Pic disk, the northeast and southwest disk midplanes are traced to
1835 AU and 1450 AU, respectively, giving a ratio of 1.27  \citep{larwood01}.  In the
case of HD 15115 the corresponding ratio is $>$1.75.  This ratio is a lower limit given that the 550 AU extent
of the west midplane is limited only by our field of view.  

A single stellar flyby, or a periodic flyby by a bound companion on an eccentric
orbit, has been studied as a potential mechanism for producing
 $\beta$ Pic's large-scale asymmetry \citep{larwood01}.  However, in a kinematic study
 of $Hipparcos$-detected stars with published radial velocities, \citet{kalas01}
did not find any perturbers that approached closer than 0.6 pc of $\beta$ Pic, though
the sample was estimated as only 20\% complete.
In the case of HD 15115, \citet{moor06} noted that another $\beta$ Pic
moving group member, HIP 12545, is located relatively nearby in sky position. 

Table 1 summarizes the observed properties of both stars. 
Their projected separation is 0.51$\degr$, which
translates to 0.38 pc at a mean heliocentric distance of 43 pc.  
Within the uncertainties, the proper motion vectors, the ($U,V$) galactic space motions,
and heliocentric distance are identical.  Furthermore,
the eastward location of HIP 12545 is in the direction of the truncated 
side of the HD 15115 debris disk.  This geometry is consistent with 
the dynamical simulation of a disk disrupted by a stellar flyby in \citet{larwood01}.
Specifically, in their Fig. 18, the long end of a highly perturbed disk is located
in the direction of periastron.  The perturber follows a parabolic trajectory such
that in a later epoch it is located in the direction opposite of periastron, or in 
the direction of the truncated side of the disk.  Periastron in these models
is $\sim$700 AU, with an initial disk radius of $\sim$500 AU.  
Overall, the ensemble of evidence favors further consideration
of HD 15115 and HIP 12545 as a possible wide-separation
multiple system with a highly eccentric orbit ($e > 0.95$).

If the heliocentric distances are in fact nearly
equivalent, then the projected sky separation
approximates the true separation.  \citet{kalas01} discuss the
Roche radius, $a_t$, of a star as containing the volume within which
the stellar potential dominates the Galactic tidal field.  Using
their Eq. 2 and the stellar mass estimates in our Table 1, 
we find $a_t$ = 1.1 pc and $a_t $= 0.7 pc for HD 15115 and HIP 12545,
respectively.   Therefore, for a small body gravitationally bound to
HD 15115, the potential well of HIP 12545 exerts a more significant 
perturbing force than the Galactic tidal field at the current epoch.  
This is not the case if we
take the $Hipparcos$ parallaxes at face value.  These give a line-of-sight
separation between the stars of $\sim$4 pc, and we derive
a 3-D separation of 5.1$\pm$2.8 pc.  We further calculate that
closest approach will occur $\sim$1 Myr in the $future$.  Therefore, improving
the parallax measurements for both stars is a critical task for
future work that would examine their possible physical association.

A prediction of the \citet{larwood01} model is that the perturber may capture
disk material, and display a tenuous and highly asymmetric tail of escaping
material pointing away from the mother disk.  To test the hypothesis that 
HD 15115 suffered a close encounter with HIP 12545,  high-contrast observations of HIP 12545
should reveal circumstellar nebulosity due to captured material.  
Since this is captured material, the nebulosity may not resemble a disk, and any tail should
point away from the mother disk (eastward).  

To futher investigate this hypothesis, we examined ACS/HRC
coronagraphic observations of HIP 12545 obtained by program GO-10487
(Principal Investigator David Ardila).   The observing technique is similar
to that described here for HD 15115.  After PSF subtraction, we do not
detect nebulosity in the vicinity of HIP 12545.  Therefore the possibility
that the extreme disk asymmetry of HD 15115 is created by dynamical
interactions with HIP 12545 does not have further supporting evidence
at the present time.

Finally, we note that among the four debris disks imaged in scattered
light in the BPMG, the dust appears depleted for HD 15115.  
The values of dust optical depth in 10$^{-4}$ units
are given as 24.3$\pm$1.1, 4.9$\pm$0.4,
29.3$\pm$1.6 and 4.0$\pm$0.3 for $\beta$ Pic (A5V),
HD 15115 (F2V), HD 181327 (F5.5V) and AU Mic (M2V), respectively \citep{moor06}.
The factor of $\sim$five smaller optical depth for HD 15115 compared
to $\beta$ Pic and HD 181327 suggests a different evolutionary
path for the disk.  Though a stellar flyby is one possibility, migration and dynamical
instabilities within a hypothetical planetary system may also
play a role in the rapid diminution of dust parent bodies around
HD 15115 \citep[e.g.][]{morbidelli97}.

\section{Summary}

Optical and near-infrared coronagraphic images of the F2 star
HD 15115 reveal a highly asymmetric debris disk with an edge-on
orientation.  We describe the morphological and photometric
properties of the disk, deferring a detailed model of scattering and
thermal emission of grains to future work. The blue scattered
light color may indicate grain properties most similar to those of the 
AU Mic debris disk, where $\Sigma_V - \Sigma_H \approx -1$ mag arcsec$^{-2}$
relative to the star  \citep{fitz07},
and less like those of $\beta$ Pic, which is predominantly red scattering
\citep{goli06}.  A key follow-up measurement would be
polarization, which in the case of AU Mic revealed highly porous
macroscopic grains \citep{graham07}.

With outer optical radius $>$550 AU, HD 15115 possesses the
second largest debris disk next to $\beta$ Pic.   However, the length asymmetry
between its west and east midplanes greatly exceeds that of $\beta$ Pic
and other disks.  HD 15115 is now the fourth debris disk discovered in
scattered light among the $\beta$ Pic moving group members.  Future work should test
our hypothesis that extreme asymmetries are due to dynamical perturbations
from the nearby M star HIP 12545. 
\acknowledgements
{\bf Acknowledgements:}  Support for GO-10896 was provided by NASA through a grant from STScI
under NASA contract NAS5-26555.

\clearpage

\begin{figure}
\plotone{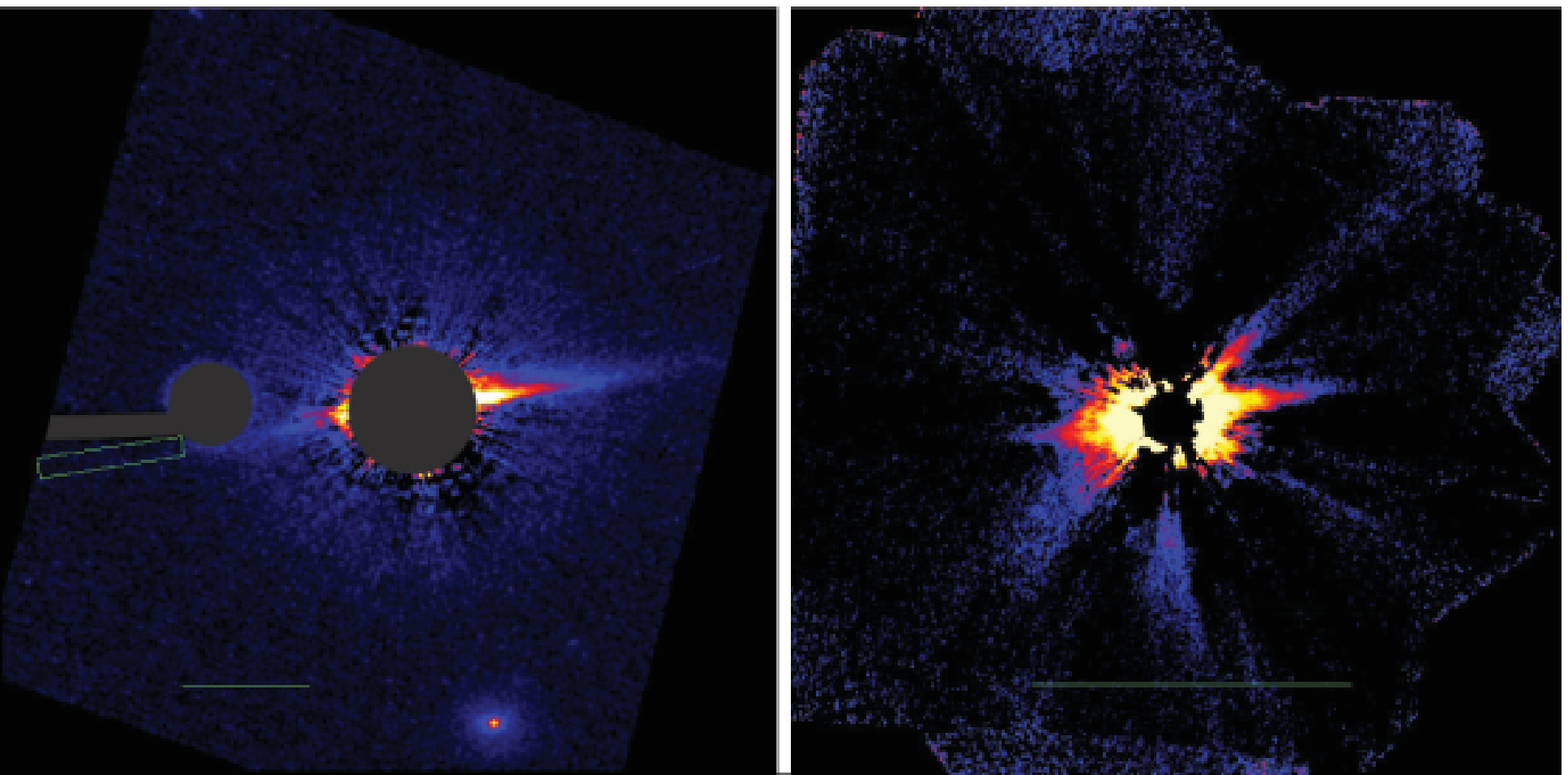}
\epsscale{0.2}
\caption{
False-color, log scale images of the HD 15115 disk as originally
discovered using the ACS/HRC F606W ($\lambda_c$= 591 nm, $\Delta\lambda$ = 234 nm)
[LEFT] and confirmed in H-band with Keck II adaptive optics [RIGHT; 2006 October 26 data].  
North is up, east is left and the scale bars span 5$\arcsec$.  
In the HST image we use gray fields over the occulting bar and 3.0$\arcsec$
occulting spot located to the left of HD 15115, as well as a gray disk covering PSF-subtraction
artifacts surrounding HD 15115 itself.   If the HD 15115 disk were a symmetric structure, then
the east side of the disk would have been detected within the rectangular box, shown below
the ACS/HRC occulting finger.  The NIR data [RIGHT] show a more symmetric disk within
2$\arcsec$ radius, with asymmetry becoming more apparent beyond 2$\arcsec$ radius.
Due to poor observing conditions, the field is contaminated by residual noise due to the
diffraction pattern of the telescope (e.g. at 2 o'clock and 7 o'clock relative to HD 15115).
However, whereas the residual diffraction pattern noise of the telescope rotates relative to the sky 
orientation over a series of exposures, the image of the disk remains fixed and it is
confirmed as real.  
 \label{fig1}}
\end{figure}

\begin{figure}
\epsscale{0.5}
\plotone{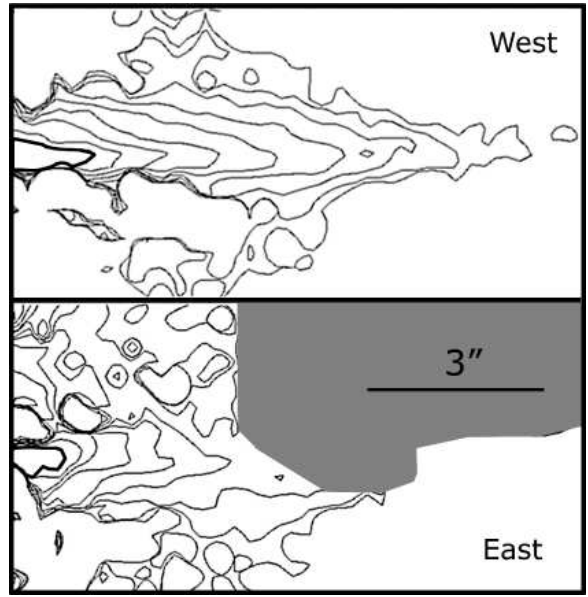}
\caption{
Surface brightness isocontours for the HD 15115 debris disk converted from F606W to
Johnson V-band (derived using STSDAS/SYNPHOT with a Kurucz model atmosphere and
the appropriate observatory parameters). 
The disk was rotate by 8$\degr$ clockwise such that the midplane
lies along a horizontal line.  The bottom
frame is the east extension, transposed across the vertical axis, and the gray region 
marks the area occupied by the ACS/HRC occulting finger and 3.0$\arcsec$ occulting spot.
The left edge of the frame corresponds to 2$\arcsec$ radius from
the star.  The innermost contour (bold) is 19.0 mag arcsec$^{-2}$
and the outermost contour represents 23.0 mag arcsec$^{-2}$,
with a contour interval of 0.5 mag arcsec$^{-2}$.
 \label{fig2}}
\end{figure}

\begin{figure}
\epsscale{0.5}
\plotone{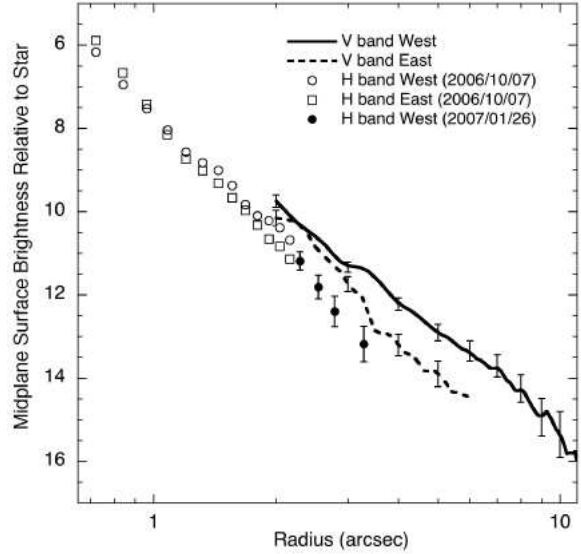}
\caption{
Radial surface brightness (mag arcsec$^{-2}$) distribution along the west and east midplanes of HD 15115.
We plot the difference between the measured disk surface brightness and the stellar
magnitudes of H=5.86 and V=6.80.  
Disk photometry was extracted from boxes 0.25$\arcsec \times 0.25\arcsec$ centered
on the midplane.  We plot a representative sample of error bars that gives the
standard deviation of the background residuals as a function of radius. 
The aperture corrections derived from point source photometry
are 0.48 and 0.57 for the H-band data in the 2006 October and 2007 January
observations, respectively, and 0.70 for the V band data.  
The H-band radial profiles between 0.7$\arcsec$ and 2.3$\arcsec$ radius
may be described by power-laws with indices
-3.7 and -4.4 for the west and east disk extensions, respectively.
In the V band, the east midplane profile may be fit by a power-law
with index -4.0 between 2.0$\arcsec$ and 6.0$\arcsec$ radius.
Thus, our data do not show a significant color gradient as a function
of radius for the east ansa.  The west midplane profile in V band may be fit by a single
power-law with index -3.0 between 2.0$\arcsec$ and 10.0$\arcsec$ radius.  This
profile is significantly shallower than the H band profile, resulting in a blue
color gradient as a function of radius for the west ansa.
\label{fig3} }
\end{figure}

\clearpage

\begin{deluxetable}{lllll}
\tabletypesize{\scriptsize}
\tablecaption{Stellar Properties\label{tbl-1}}
\tablewidth{0pt}
\tablehead{
\colhead{} & \colhead{HD 15115} & \colhead{HIP 12545} & \colhead{Ref.}
}
\startdata
Spectral Type 			& F2				& M0 & Hipparcos\\
$m_V$ (mag)			& 6.79			& 10.28 & Hipparcos\\
Mass (M$_\odot$)		&1.6				& 0.5	& Astrophys. Quant.\\
Distance (pc)  			& $44.78^{+2.22}_{-2.01}$ 	& $40.54^{+4.38}_{-3.61}$ &Hipparcos\\
RA	(ICRS)			& 02 26 16.2447	& 02 41 25.89 & Hipparcos\\
DEC	 (ICRS)			& +06 17 33.188	& +05 59 18.41&Hipparcos \\
$\mu_\alpha$(mas/yr)	&$86.09 \pm 1.09$	& $82.32 \pm 4.46$ &Hipparcos \\
$\mu_\delta$ (mas/yr)	&$-50.13 \pm 0.71$	& $-55.13 \pm 2.45$ &Hipparcos\\
$\mu_\alpha$(mas/yr)	&$87.1 \pm 1.2$	&  $82.3 \pm 4.3$ &Tycho-2\\
$\mu_\delta$ (mas/yr)	&$-50.9 \pm 1.2$	& $-55.1 \pm 2.7$ &Tycho-2\\
U (km / s)				&$-13.2\pm1.9$ 	&    $-14.0 \pm 0.5$& a  \\
V (km / s)				&$-17.8 \pm 1.2$ 	&	$-16.7 \pm 0.9$& a\\
W (km / s)				&$-6.0 \pm 2.3$  	& $-10.0 \pm 0.5$& a \\
\enddata
\tablenotetext{a}{Galactic kinematics for HD 15115 and HIP 12545 from
\citet{moor06} and \citet{song03}, respectively
}
\end{deluxetable}
\clearpage

\end{document}